\documentclass[12pt,a4paper]{article}
\usepackage[T2A]{fontenc}
\usepackage[cp1251]{inputenc}
\usepackage{amsmath,amssymb,amsthm,amsfonts,amscd}
\begin{document}
\newtheorem{lemma}{�����}
\newtheorem{proposition}{�����������}
\newtheorem{example}{������}
\newtheorem{remark}{���������}
\newtheorem{theorem}{Theorem}
\newtheorem{corollary}{���������}

\def\Z{{\Bbb Z}}
\def\R{{\Bbb R}}
\def\RP{{\Bbb R}\!{\rm P}}
\def\N{{\Bbb N}}
\def\C{{\Bbb C}}
\def\A{{\bf A}}
\def\D{{\bf D}}
\def\k{{\bf k}}
\def\E{{\bf E}}
\def\F{{\bf F}}
\def\V{\vec{\bf V}}
\def\L{{\bf L}}
\def\M{{\bf M}}
\def\c{{\bf c}}

\def\fr{{\operatorname{fr}}}
\def\st{{\operatorname{st}}}
\def\mod{{\operatorname{mod}\,}}
\def\cyl{{\operatorname{cyl}}}
\def\dist{{\operatorname{dist}}}
\def\grad{{\bf{grad}}}
\def\div{{\operatorname{div}}}
\def\rot{{\operatorname{rot}}}

\def\R{{\Bbb R}}
\def\B{{\bf B}}
\def\e{{\bf e}}
\def\L{{\bf L}}
\def\valpha{\vec{\alpha}}
\def\vxi{\vec{\xi}}
\sloppy
\title{Quadratic  helicities and the energy of magnetic fields}
\author{P.M.Akhmet'ev}
\date{}
\maketitle

\begin{abstract}
Two non-local asymptotic invariants of magnetic fields for the ideal magnetohydrodynamics are introduced. 
The velocity of variation of the invariants for a non-ideal magnetohydrodynamics with a small magnetic dissipation is estimated.
By means the invariants spectra of electromagnetic fields are investigated. A possible role of higher magnetic helicities 
during a relaxation of magnetic fields is discussed.
\end{abstract}
\[  \]
Key words: magnetic helicity, asymptotic Hopf invariant, Gauss integral, magnetic energy, induction equation, linking number.
\[  \]
Codes: MSC: 35Qxx, 57Mxx, 76Fxx.

\section{Introduction}

The magnetic helicity invariant plays a pivotal role for investigation in magnetohydrodynamics, see
[A-Kh],[Z-R-S],[Ro]. A topological meaning of the helicity invariant, which is also called the asymptotic Hopf invariant and is denoted by $\chi$
was clarified by V.I.Arnol'd using an asymptotic limit, when the length of magnetic lines tends to $+\infty$. 
In the present paper from this point of view a two new simplest higher asymptotic invariants of magnetic fields, which are called the quadratic helicities, are introduced.   

These two invariants are denoted by $\chi^{[2]}$, $\chi^{(2)}$, these invariant are non-negative. Strictly speaking, the square of the magnetic helicity $\chi^2$ 
determines the third example of quadratic helicities, unlike the first two considered invariants $\chi^2$ is a local invariant. 
Quadratic helicities are defined for an arbitrary magnetic field with a compact domain in $\R^3$. We will considered also magnetic fields on the standard sphere and on the standard 3-torus. 

In Theorem  $\ref{th1}$ is proved that the quadratic helicities are invariants in the ideal magnetohydrodynamics, i.e. are preserved by the
one-parameter families of volume-preserved diffeomorphisms. 
In the case when the magnetic field is presented by a finite set of magnetic tubes (see the definition in [B-F]), the topological meaning of the helicity is simple and is considered in Examples 1, 2.

 Correlation tensors 
 $\delta^{[2]}$, $\delta^{(2)}$, which are determined an upper bound for the invariants $\chi^{[2]}$, $\chi^{(2)}$ are constructed.
 This correlation tensors are immediately calculated using spectral decompositions of the magnetic field.
 The upper and lower estimations, which are given by the formula 
 $(\ref{11})$, $(\ref{12})$,  gives answers to a question by V.I.Arnol'd (2008). 

In section 3 applications for the induction equation are investigated. The induction equation describes the magnetic field in a non-ideal 
liquid conductive medium, assuming that the velocity of the medium is known. It is proved that the velocity of the variations of the
quadratic helicity $\chi^{(2)}$ is estimated using magnetic and velocity fields, which are smooth. A force-free magnetic fields
on the standard 3-torus are investigated. 

In section 4 the results of section 2 are investigated from point of view the ergodic Birkhoff theorem. The quadratic helicity $\chi^{(2)}$ 
admits the following meaning: this is the dispersion of the fluctuation of the helicity density over the space of magnetic lines.   

In section 5 general facts, which are related with a reconnection of magnetic fields and with a relaxation of magnetic fields in the sense of
J.B.Taylor are discussed. A possible role of higher magnetic helicities and  of quadratic magnetic helicity is mentioned.  

In section 6 simple calculations of the spectrum of the correlation tensor 
$\delta^{(2)}$ for magnetic fields on the 3-torus is investigated. 
Assuming that the magnetic field is described by  a polynomial spectrum (this assumption is acceptable in the most applications),
we prove that the spectrum of the quadratic helisity $\chi^{(2)}$ is between the spectrum of the square of the magnetic energy and 
of the spectrum of the square of the magnetic helisity. 

The author is grateful to B.A. Borisov, V.B.Semikoz, D.D.Sokoloff for discussions.

\section{Definitions and the main inequalities}

Let us assume that a magnetic field
 $\B$, $\div(\B)=0$, are inside the ball $D$ of the radius $r$. 
 Let us define the helicity $\chi$ of the magnetic field $\B$  by the formula: 
$$
\chi =  \int (\A,\B) dD,
$$
where $\A$ is the vector-potential of $\B$, namely, the vector, which is satisfies the following conditions: 
$$\rot(\A)=\B, \quad \div(\A)=0.$$ 
In the case $D \subset \R^3$ the boundary condition for $\A$ is the following:  $\A(x) \to 0$, if $x \to
\infty$. (In some problem one can assume that  $\A(x)$ is tangent to  $\partial D$.)
The magnetic helicity $\chi$ is an invariant in the case, when $\B$ is frozen-in,  which is related with the magnetic energy  
$$U=\int (\B,\B)dD.$$

\subsection*{Asymptotic Hopf invariant}

Let us define a Gaussian  linking coefficient of two trajectories of the magnetic field $\B$, which are issued from points 
 $x_1, x_2$ by the time $T$ as follows:

\begin{eqnarray}\label{Lambda}
\Lambda(T;x_1,x_2) = \frac{1}{4 \pi T^2}
\int_{0}^{T} \int_{0}^{T} \frac{\langle
\dot{x}_1(\tau_1),\dot{x}_2(\tau_2),x_1(\tau_1)-x_2(\tau_2) \rangle}{ \|
x_1(\tau_1) - x_2(\tau_2)\|^3} d\tau_1 d\tau_2,
\end{eqnarray}
where $x_i(\tau_i) = g^{\tau_i}(x_i)$, $i=1,2$ are the trajectory of the point $x_i$, which is given by the flow of the magnetic field
$\B$, 
 $\dot{x}_i(\tau_i) = \frac{d}{d\tau_i} g^{\tau_i}x_i$ are the corresponding velocity vectors.

For an arbitrary $T>0$ the following formula is satisfied:
$$
\chi = \int \int \Lambda(T;x_1,x_2) dx_1 dx_2.
$$

\subsubsection*{A non-formal definition of the helicity invariant $\chi$ using asymptotic Hopf invariant}
Define
$$
\chi = \int \int \Lambda(l_1,l_2) d\Omega d\Omega,
$$
where $\Omega$ is the spectra of the magnetic field $\B$, by this the space of all magnetic lines  $l_1,l_2 \in \Omega$,
$\Lambda(l_1,l_2)$ is the (asymptotic) linking number of the pair of magnetic lines.

\subsubsection*{A non-formal definition of the quadratic helicities $\chi^{[2]}$, $\chi^{(2)}$}
Define
$$\chi^{[2]} = \int  P^{[2]}(l_1,l_2) d\Omega $$
$$\chi^{(2)} = \int (\int  \Lambda(l_1,l_2) d\Omega)^2 d\Omega = \int \int \int P^{(2)}(l_1,l_2,l_3) d\Omega d\Omega d\Omega, $$
where $P^{[2]}(l_1,l_2)= \Lambda^2(l_1,l_2)$, $P^{(2)}(l_1,l_2,l_3) =
\frac{1}{2} \Lambda_{\B}(l_1,l_2)\Lambda_{\B}(l_3,l_1)$,
$l_1,l_2,l_3 \in \Omega$.

\begin{remark}\label{rem}
In the previous formula the 
polynomial $P$ could be replaced by an arbitrary (symmetric) polynomial of 
the $\frac{k(k-1)}{2}$ variables $\Lambda(l_i,l_j)$, $0 \le i < j \le k$. A more interesting example in the case $k=3$ is considered in 
$[Akh]$: the invariant  is defined as the integral of a finite-order (asymptotic) invariant of all triples 
of magnetic lines, which is not expressed from the pairwise linking numbers of the components
(see discussion in section 5).
\end{remark}

Let us start by the precise definitions.

\subsubsection*{Definition of the quadratic helicity $\chi^{(2)}$}

Define the asymptotic quadratic linking number $\Lambda^{(2)}$ by the formula: 
\begin{eqnarray}\label{Lambda2}
\Lambda^{(2)}(T;x) = \frac{1}{T^2}
(\int_{0}^{T}  (\dot{x}(\tau),\A)d\tau)^2,
\end{eqnarray}
where $x(\tau) = g^{\tau}(x)$ is the trajectory of the point $x$, which is given by the flow of $\B$,
 $\dot{x}(\tau) = \frac{d}{d\tau} g^{\tau}x$ is the corresponding velocity vector.

 Define  $\chi^{(2)}$ by the formula: 
\begin{eqnarray}\label{(2)}
\chi^{(2)} = \limsup_{T \to +\infty} \int \Lambda^{(2)}
(T;x)dD.  
\end{eqnarray}

\subsubsection*{Definition of the quadratic helicity  $\chi^{[2]}$}

Define the asymptotic quadratic linking number $\Lambda^{[2]}$ by the formula: 
$$
\Lambda^{[2]}(T;x_1,x_2) = \frac{1}{T^4}
\int_{0}^{T} \int_0^T (\dot{x_1}(\tau_1),\A(x_2(\tau_2))^2d\tau_1 d\tau_2.
$$

Define $\chi^{[2]}$ by the formula: 
$$ \chi^{[2]} = \limsup_{T \to +\infty} \int \Lambda^{[2]}
(T;x_1,x_2)dD.  $$

\subsubsection*{The invariant $\chi^{(2)}$ is well-defined}

By the Cauchy�Bunyakovsky�Schwarz inequality we get: 

\begin{eqnarray}\label{noneq}
 \Lambda^{(2)}(T;x) \le \frac{1}{T} \int (\dot{x}(\tau),\A)^2 d\tau.  
\end{eqnarray} 

Therefore:
 $$ \int \Lambda^{(2)}(T;x)dD \le \frac{1}{T}\int \int (\dot{x}(\tau),\A)^2 d\tau dD = \int(\B,\A)^2dD.$$

\begin{eqnarray}\label{delta}
 \int \Lambda^{(2)}
(T;x)dD \le \delta^{(2)},
\end{eqnarray}
where

\begin{eqnarray}\label{delta2}
\delta^{(2)} = \int(\B,\A)^2dD.
\end{eqnarray}

\subsubsection*{The invariant $\chi^{[2]}$ is well-defined}

Let us prove (see the analogous formula $(\ref{delta})$) that
\begin{eqnarray}\label{delta[2]}
\chi^{[2]} \le \delta^{[2]}, 
\end{eqnarray}
where
\begin{eqnarray}\label{delta[]}
\delta^{[2]} = \int \int \delta^{[2]}(\B(x_1),\B(x_2)) dx_1 dx_2, 
\end{eqnarray}
$$ \delta^{[2]}(\B(x_1),\B(x_2)) = (\B(x_1), \A(x_2;x_1))^2=(\A(x_1;x_2),\B(x_2))^2.$$

The integral
 $(\ref{delta[]})$ is convergence, because 
$\delta^{[2]}(\B(x_1),\B(x_2)) = O(dist^{-2}(x_1,x_2))$, where $dist(x_1,x_2) \to 0$.

\subsection*{$\chi^{[2]}$, $\chi^{(2)}$ are (non-local) invariants of frozen-in magnetic fields}

Let us prove that the quadratic helicities 
$\chi^{[2]}$, $\chi^{(2)}$ are invariants with respect to volume-preserved diffeomorphisms.

\subsubsection*{$\chi^{(2)}$ is an invariant}

Using induction equation  (see. [M], eq. (2.118)) in the ideal case we get: 
$$\frac{\partial \A}{\partial t} = \bf{v} \times \B - \grad f,$$
where $f$ is a function in the domain $U$ with the right boundaries conditions at the infinity, which satisfies the equation
$$ \Delta f = \div({\bf{v}} \times \B).$$ 
This equality will be used in the form, which is proposed in  [Ma].
 
A magnetic line $x(\tau)$ of $\B$ is transformed as follows: 
 $x'(\tau) = x(\tau) + dx(\tau) = x(\tau) + \rot({\bf{v}}(x(\tau)) \times \B(x(\tau)))$.

At each point
$x(\tau)$ of the trajectory we get:
$$(\frac{\partial}{\partial t} + L_{\bf{v}})\A = \grad f,$$
$$(\frac{\partial}{\partial t} + L_{\bf{v}})\B = 0, $$ 
where by 
$L_{\bf{v}}$ the Lee derivative with respect to the velocity of the domain is denoted.

Recall that by definition at each point  $x(\tau)$ of the trajectory we get
$\dot{x}(\tau) = \B(x(\tau))$. We shall use this denotation, by the velocity vector we shall means the vector $\bf{v}$

The integral 
$\Lambda^{(2)}$ is transformed as following: 
$$ \Lambda^{(2)}(T;x) \mapsto \Lambda^{(2)}(T;x) + $$
$$\frac{2}{T^2}
(\int_{0}^{T}  (\B(x(\tau)),\A(x(\tau)))d\tau)(\int_{0}^{T}  (\dot{x}(\tau),(\frac{\partial}{\partial t} +
 L_{\bf{v}})\A(x(\tau))d\tau \quad + $$
 $$
 \int_{0}^{T}  ((\frac{\partial}{\partial t} + L_{\bf{v}})\B(x(\tau)),\A(x(\tau)))d\tau).$$
Therefore, we get: 
$$ \Lambda^{(2)}(T;x) \mapsto \Lambda^{(2)}(T;x) + \int_{0}^{T}  (\B(x(\tau)), \grad f(x(\tau)))dt. $$
To prove that $\chi^{(2)}$ is an invariant it is sufficiently to prove that the transformation 
$$\int \Lambda^{(2)}(T;x)dD \mapsto \int \Lambda^{(2)}(T;x)dD + $$
$$\frac{2}{T^2}
(\int \int_{0}^{T}  (\B(x(\tau)),\A(x(\tau)))dt dD)(\int \int_0^T (\B(x(\tau)),\grad f(x(\tau))) dt dD) + $$
$$ \frac{1}{T^2} (\int \int_0^T (\B(x(\tau)),\grad f(x(\tau)) dt dD)^2 $$
is the identity if $T \to + \infty$.

By the Newton-Leibniz formula we get:
$$ \int_0^T (\B(x(\tau)),\grad f(x(\tau)))dD = f(x(o))-f(x(T)) \le C, $$
where $C$ depends of $f$, but not of $T$.
Therefore we get: 
$$\int \Lambda^{(2)}(T;x)dD \mapsto \int \Lambda^{(2)}(T;x)dD + T^{-1}C_1,$$
where $C_1$ is bounded for $T \to +\infty$. Therefore we get: 
$$ \limsup_{T \to +\infty}  \int \Lambda^{(2)}_{\B}(T;x)dD\mapsto \limsup_{T \to +\infty} \int \Lambda^{(2)}(T;x)dD.$$
It is proved that $\chi^{(2)}$ is an invariant of volume-preserved diffeomorphis of the domain $D$.

\subsubsection*{$\chi^{[2]}$ is an invariant}

The integral 
$\Lambda^{[2]}$ is transformed as follows:  
\begin{eqnarray}\label{cal}
 \Lambda^{[2]}(T;x_1,x_2) \mapsto \Lambda^{[2]}(T;x_1,x_2) + 
 \end{eqnarray}
$$\frac{2}{T^2}\int \int(\B(x_1(\tau_1)),\A(x_2(\tau_2));x_1(\tau_1))d\tau_1 d\tau_2 \cdot$$
$$\cdot(\int \int (\B(x_1(\tau_1)),\grad \phi(x_2(\tau_2);x_1(\tau_1))d \tau_1d\tau_2 + $$
$$ \int(\psi(x_2(0);x_1(\tau_1)))+(\psi(x_2(0);x_1(T)))d \tau_1),$$
where $\phi(x_2(\tau_2);x)$ $\phi(x_1(\tau_1);x)$ are the corresponding family of smooth functions of the variable $x$,
$\psi(x_2(0);y)$, $\psi(x_2(T);y)$, $\psi(x_1(0);y)$, $\psi(x_1(T);y)$ are family of smooth functions of the variable $y$.
Each function 
$\phi(x_2(\tau_2);x)$ (correspondingly, each function $\phi(x_1(\tau_1);x)$) of the considered  $\tau_2$--family
(correspondingly,  $\tau_1$--family) has a singularity at the point  $x=x_2(\tau_2)$ 
(correspondingly, at the point $x=x_1(\tau_1)$) of the order not more then $-1$, the coefficient at the singular point 
is calculated from the initial data 
(from the magnetic and the velocity field) and depends no of $T$. 
Each function 
$\psi(x_2(0);y)$, $\psi(x_2(T);y)$, $\psi(x_1(0);y)$, $\psi(x_1(T);y)$
has the singularity at the point $y=x_2(0)$ (correspondingly, at the point  $y=x_2(T)$,  $y=x_1(0)$, $y=x_1(T)$) 
of the order not more then $-1$, and the coefficient of the singular point is also calculated fro the initial data.

The magnetic field  
$\B(x_2(\tau_2))$ along the line $x_2(\tau_2)$ is represented by the family of the dipoles with the axis along the magnetic line.
In the case when the line is closed, the integral of the considered family of dipoles converges outside the magnetic line and is equal to
the ill vector-field along the magnetic line.  

In the case the magnetic line is non-closed, the integral of the considered family is represented by the sum of the ill vector-field, which is considered above, and the potential vector field, which has the singularities at the end-points of the order $-2$. 
In a one-parameter family of volume-preserved of diffeomorphisms the one-parameter family $\phi(x_2(\tau_2);y)$ is defined as
the gauge of the scalar potential along the magnetic line $x_1(\tau_1)$. 
The family of functions  
$\psi(x_1(\tau_1);x_2(0))$, $\psi(x_1(\tau_1);x_2(0))$ 
are appeared as the scalar product of the gauge of the vector-potential of the magnetic field
with singularities at the points 
 $x_2(0), x_2(T)$ with the magnetic field along the magnetic line. 
The formula $(\ref{cal})$ is proved.

The integral 
\begin{eqnarray}\label{int1}
\int (\B(x_1(\tau_1),\grad \phi(x_2(\tau_2);x_1(\tau_1))d\tau_1  
\end{eqnarray}
is estimated by 
$C_1(\vert\vert x_1(0)-x_2(\tau_2)\vert\vert^{-1} + \vert\vert x_1(T)-x_2(\tau_2)\vert\vert^{-1})$ 
for a suitable constant  $C_1$, which depends no of $T$.
The  functions
\begin{eqnarray}\label{int2}
\int (\psi(x_1(\tau_1);x_2(0))d\tau_1, \cdot  \int (\psi(x_1(\tau_1);x_2(T))d\tau_1  
\end{eqnarray}
are estimated as
$C_2(\vert\vert x_1(\tau_1)-x_2(0)\vert\vert^{-1} + \vert\vert x_1(\tau_1)-x_2(T)\vert\vert^{-1})$ 
for a suitable constant $C_2$, which depend no of $T$.

It is proved that the gauge term in the integral
$(\ref{cal})$ is estimated over the variables $\tau_1, \tau_2$ by 
\begin{eqnarray}\label{call}
T^{-1}C_3  \int \int \vert \vert x_1(\tau_1) - x_2(\tau_2) \vert \vert^{-2} d \tau_1 d \tau_2, 
\end{eqnarray} 
for a suitable constant $C_3$, which depends no of  $T$.
After the integration over the space we get
$$ \limsup_{T \to +\infty}  \int \Lambda^{[2]}_{\B}(T;x)dD\mapsto \limsup_{T \to +\infty} \int \Lambda^{[2]}(T;x)dD.$$
It is proved that 
$\chi^{[2]}$ is invariant with respect to one-parameter families of volume-preserved diffeomorphisms.

\subsection*{Inequalities between the quadratic helicities}

Let us prove the following result:

\begin{theorem}\label{th1}

The following inequalities are satisfied:

\begin{eqnarray}\label{11}
\delta^{(2)} \ge \chi^{(2)},
\end{eqnarray}
\begin{eqnarray}\label{12}
\delta^{[2]} \ge \chi^{[2]},
\end{eqnarray}
\begin{eqnarray}\label{13}
\frac{\delta^{[2]}}{Vol(D)} \ge \delta^{(2)},
\end{eqnarray}
\begin{eqnarray}\label{14}
\chi^{[2]} \ge \frac{2\chi^{(2)}}{Vol(D)} \ge
\frac{2\chi^2}{Vol^2(D)} \ge 0.
\end{eqnarray}
All the values, which are included in the sequence 
 $(\ref{14})$, have the dimension ��$^4$��$^{2}$.
\end{theorem}

\subsubsection*{Proof of Theorem  $\ref{th1}$}

The upper bounds 
$(\ref{11})$, $(\ref{12})$ are following from the definition of $\chi^{(2)}$, see the inequalities  $(\ref{delta})$, $(\ref{delta[2]})$.
 
 The estimation
$ Vol(D)\delta^{[2]} \ge \delta^{(2)}$ is proved by the Cauchy�Bunyakovsky�Schwarz inequality:
$$ \int \int (\B(x_1),\A(x_2;x_1))^2 dx_1 dx_2 \le \int Vol(D)(\int (\B(x_1),\A(x_2;x_1)) dx_2)^2 dx_1 .$$ 

Let us prove the following inequality:
$2 Vol(D)\chi^{[2]} \ge \chi^{(2)}$.  Using the inequality
$2\Lambda(x_1,x_2;T)\Lambda(x_1,x_3;T) \le  \Lambda^2(x_1,x_2;T) + \Lambda^2(x_1,x_3;T)$,
where $x_i \in l_i$, $i=1,2,3$, by the integration over $x_2$, $x_3$ we get: 
$$2 \Lambda^{(2)}(x_1;T) \le Vol(D)\int \Lambda^2(x_1,x_2;T)dx_2 + Vol(D) \int\Lambda^2(x_1,x_3;T)dx_3. $$ 
After the integration over 
$x_1$ we get the required inequality. 
 
Let us prove the following inequality: 
$ Vol(D)\chi^{(2)} \ge \chi^2. $
Consider the function 
$\Lambda^{(2)}(T;x)$ over $D$, which is defined by the formula 
$(\ref{Lambda})$. Let us define the function  $\Lambda_{\A}(T;x)$ by the formula: 
$$\Lambda_{\A}(T;x) = \frac{1}{T}
\int_{0}^{T}  (\B(x(\tau),\A(x(\tau))d\tau.$$
Evidently, we get
$\Lambda_{\A}^2(T;x)= \Lambda^{(2)}(T;x)$.

From the Cauchy�Bunyakovsky�Schwarz inequality we get:
$$ Vol(D) \int \Lambda^{(2)}(T;x) dD \ge (\int \Lambda_{\A}(T;x) dD)^2 .$$
Let us recall the 
 $\int \Lambda_{\A}(T;x)dD = \chi$ depends no of  $T$. Take the upper limit in the previous inequality.
Theorem   
 $\ref{th1}$ is proved.

\subsection*{Topological meaning of the quadratic helicities}

\subsubsection*{Example 1}

Assume that the magnetic field
$\B$ is localized into the only magnetic plate tube  $L \subset D$, 
inside the tube all trajectories are closed. Assume that the magnetic tube is characterized by the following parameters:
\[  \]

-- $\Phi$ is the magnetic flow thought the cross-section of the tube, 

-- $\kappa \in \Z$ is the twisting coefficient along the central axis 
(for the plate magnetic tube this coefficient is equal to the linking number of the pair of magnetic lines), 

-- $L$ is the length of the central line of the tube, 

--$Vol$ is the volume of the magnetic tube. 
\[  \]

The magnetic energy is given by:
$$U = \Phi^2 L,$$

The magnetic helicity is given by: 
$$\chi = \kappa \Phi,$$

The quadratic helicities are given by:
$$\chi^{(2)} = \frac{\kappa^2 \Phi^2}{Vol},$$
$$\chi^{[2]} = \frac{\kappa^2 \Phi^2}{2Vol^2}.$$
\[  \]

Let us consider the limit (the cross-section of the magnetic tube tends to zero)
$\kappa = const$, $\Phi = const$, $L=const$, $Vol(L) \to 0$.
 \[  \]
 Therefore:
$$ U = const, \quad \chi  = const, \quad \chi^{(2)} \to +\infty,$$
$$ \chi^{[2]} \to + \infty, \quad \chi^{[2]} = O(\chi^{(2)}).$$

\subsubsection*{Example 2}

Assume that the magnetic field $\B$ is localized into two plate untwisted magnetic tubes  $L_1, L_2 \subset D$, 
inside the each tube, inside the each tube magnetic lines are closed. 
Then
$$\chi^{[2]} = Vol^{-1}(L_1)Vol^{-1}(L_2)\chi^2,$$ 
$$\chi^{(2)}=(Vol(L_1)+Vol(L_2))^{-1}\chi^2.$$
For the considered configuration of the magnetic tubes the magnetic energy
 $U$ cannot be an arbitrary small.

\section{Application to the induction equation}

The following equation describes the magnetic field in the liquid medium, assuming that the spacial-time distribution of the velocity
${\bf{v}}$ is given:  

\begin{eqnarray}\label{ind}
 \frac{\partial \B}{\partial t} = \rot({\bf{v}} \times \B) + \alpha \rot \B - \eta \rot \rot \B. 
\end{eqnarray}

The second term at the right side of the equation 
 $(\ref{ind})$ leads to growth of the mean magnetic fields, this term is due to hydrodynamic
(see. [Z-R-S] p. 146, equation  (9); [Ro]) or, by quantum effects [S-S], which are due to an asymmetry of the neutrino particle.
The third term at the right side of the equation 
 $(\ref{ind})$ leads to relaxation of the mean magnetic fields, this terms is related with the dissipation of the magnetic field
[M, eq. (2.118)], [Z-R-S, p. 146, equation (9)].

It is naturally to assume that the investigation of the quadratic helicities 
$\chi^{[2]}$, $\chi^{(2)}$ in the framework of the induction equation with the $\alpha$-term
is interesting, because solutions of this equation are not invariant and are not skew-invariant with respect to the mirror symmetry of the space.
Recall that the helicity invariant $\chi$ is skew-symmetric and the invariants
$\chi^{[2]}$, $\chi^{(2)}$ are symmetric with respect to the mirror symmetry of the magnetic field in space.

The following equality is well defined:
$$ \frac{d\chi}{dt} = -2 \eta \int (\B,\rot \B)dD + 2\alpha \int(\B,\B)dD = -2\eta \chi^c + 2\alpha U.$$
This equality is well-known, see 
[A-Kh],[S-S]. A geometric sense of this equality is presented in  [A-K-K]. Let us prove an analogous equality for the quadratic helicity $\chi^{(2)}$.

\begin{theorem}\label{th2}
Assume the induction equation 
 $(\ref{ind})$ is satisfied, then the following inequality holds: 
$$\frac{d \sqrt{\chi^{(2)}}}{dt} \le \eta \sqrt{\int (\rot \B, \B)^2 dD} + \eta \sqrt{\int (\rot \rot \B, \A)^2 dD} +$$
$$\alpha \sqrt{\int (\B,\B)^2 dD} + \alpha \sqrt{\int (\rot \B,\A)^2 dD} +$$
$$\eta (\int (\rot \rot\B,\rot \rot \B)^4 dD)^{1/8}(\int(\A,\A)^2 dD)^{1/4} + $$
$$\alpha (\int (\rot \B,\rot \B)^4 dD)^{1/8} (\int(\A,\A)^2 dD)^{1/4}. $$

\end{theorem}

\subsubsection*{Remark}
The right side 
 $\frac{d \sqrt{\chi^{(2)}}}{dt}$ of the inequality in Theorem $\ref{th2}$ is defined as the lower bound of
 the limit of the difference quotient. 
\[  \]

To prove the theorem the following lemma is required.

\begin{lemma}\label{lemma}
Assume $x(t)$, $t \in [0;T]$ is a magnetic line from a point $x(0)$ into a $x(T)$ during the time $T$. Assume that $g(t,\varepsilon)$, $0<\varepsilon \le \varepsilon_0$ is a smooth one-parameter family of the curves, such that  $g(0,\varepsilon)=x(0)$, $g(T,\varepsilon)=x(T)$ 
for an arbitrary $\varepsilon$. Then for $\varepsilon \to 0$ the following equality is satisfied: 
$$ \int_0^T(\A,\dot{g}(t,\varepsilon))dt = \int_0^T(\A,\dot{x}(t))dt + O(\varepsilon^2).$$
\end{lemma}

\subsubsection*{Proof of Lemma $\ref{lemma}$}

Let us consider a closed curve 
$\gamma=x(t) \cup g(T-t,\varepsilon)$, which is bounded a thin elongated disk $S$.
By Stokes Lemma the following equality is satisfied:
$$ \oint (\dot{\gamma},\A) d \gamma = \int \B dS, $$
where the right side of the expression is the flow of 
$\B = \rot \A$ thought the surface $S$. Evidently, that for 
$\varepsilon \to 0$ the flow $\B$ is of the order $\varepsilon^2$, because this flow is estimated by the oriented area of the projection of
the surface $S$ onto the plane of a cross-section of the magnetic line 
$x(t)$. Lemma $\ref{lemma}$ is proved. 
\[  \]

\begin{corollary}\label{cor}
Assume that in Lemma  $\ref{lemma}$ the equality $x(T)=g(T,\varepsilon)$ is satisfied up to 
$O(\varepsilon)$. Denote  
$$g(T,\varepsilon)-x(T)$$ 
by $\bf{l}$. The following equality is satisfied: 
$$ \int_0^T(\A,\dot{g}(t,\varepsilon))dt = \int_0^T(\A,\dot{x}(t))dt + (\A(x(T)),{\bf {l}}) + O(\varepsilon^2).$$
\end{corollary}

\subsubsection*{Proof of Corollary $\ref{cor}$}

Let us consider the curve 
$\gamma' = x(t) \cup g(T-t,\varepsilon)$, which is completed by the segment  $\bf{l}$ to a closed curve, denoted by
$\gamma$. The integral
$ \oint (\dot{\gamma},\A) d \gamma$ over the segment  $\bf{l}$
is estimated up to  $O(\varepsilon^2)$ by the value $(\A(x(T)),{\bf {l}})$.
The required equation is a particular case of the equation in Lemma $\ref{lemma}$. 
\[  \]

\subsubsection*{Proof of Theorem $\ref{th2}$}

Consider the following inequality 
$$\frac{d \sqrt{\chi^{(2)}}}{dt} \le \sup_T \frac{d}{dt} \left(\int \Lambda^2(T,x)dD \right)^{1/2},$$
where the both sides are considered by its absolute values.
For an arbitrary $T$ let us transform the right side of the inequality into the product of the two factors:
\begin{eqnarray}\label{KB}
 \left( \int \Lambda(T,x) \frac{d \Lambda(T,x)}{dt} dD \right) \left( \int \Lambda^2(T,x) dD   \right)^{-1/2}.
\end{eqnarray}
Let us use the equality:
\begin{eqnarray}\label{dt}
 \frac{d \Lambda(T,x)}{dt} = T^{-1}\int_0^T ((\frac{\partial}{\partial t} + L_{\bf{v}}) \B(x(\tau),\A)d\tau \quad + 
\end{eqnarray} 
$$  T^{-1}\int_0^T (\B(x(\tau),(\frac{\partial}{\partial t} + L_{\bf{v}}) \A(x(\tau)))d\tau \quad +  \quad T^{-1}\int_0^T \Lambda_{\bf{v}}(x(\tau)) d\tau. $$
The third term of this formula contains the factor
$\Lambda_{\bf{v}}(x(\tau))$, which is the partial derivative over $t$ of the function  $\Lambda(x(\tau))$ 
of the variable $\tau \in [0,T]$, which is defined by a shift of the magnetic line
$x(\tau)$ of  $\B$ into the magnetic line $x(\tau) + dx(\tau)$ of  $\B + d\B$, which is issued from the same point 
$x(0)$ at the time $t + dt$.

Let us put 
$(\ref{dt})$ into the expression $(\ref{KB})$ and let us apply to each of the three terms the the Cauchy�Bunyakovsky�Schwarz inequality.
The values $(\frac{\partial}{\partial t} + L_{\bf{v}}) \B(x(\tau))$, $(\frac{\partial}{\partial t} + L_{\bf{v}}) \A(x(\tau))$ 
are determined from the equation $(\ref{ind})$. As the result the first two terms of the equation  
$(\ref{KB})$ are transformed into the first four terms into the right side 
of the inequality.

Let us prove that the third term in 
 $(\ref{KB})$ is transformed into the last term of the inequality.  Using Corollary
$\ref{cor}$, according which the following equation is satisfied: 
 $$\int_0^T \Lambda_{\bf{v}}(x(\tau)) d\tau = (\int_0^T (\frac{\partial}{\partial t} + L_{\bf{v}})x(\tau) d\tau,\A(x(T))).$$
 Putting in this equation the value
$(\frac{\partial}{\partial t} + L_{\bf{v}})x(\tau)$ we get that the considered term is estimated by the absolute value of the scalar product:
$$ T^{-1} (\int_0^T (\alpha \rot \B(x(\tau)) - \eta \rot \rot \B(x(\tau)))d\tau,\A(x(T))),$$
where the first vector is defined by the integral over the path
 $x(\tau)$. Estimate the value of the scalar product by the product of the norms of the vectors.
 Then the previous expression is estimated by the sum of the integrals as follows:
$$ T^{-1} \alpha \int_0^T \sqrt{(\rot \B(x(\tau)))^2} \sqrt{\A^2(x(T))}d\tau \quad + $$
$$T^{-1} \eta \int_0^T \sqrt{(\rot \rot \B(x(\tau)))^2} \sqrt{\A^2(x(T))}d\tau.$$
Put this expression into the third term of the formula
$(\ref{dt})$,  and apply the Cauchy�Bunyakovsky�Schwarz inequality twice to the corresponding term of the formula  $(\ref{KB})$
by the integration over  $dD$ and $dT$. Theorem $\ref{th2}$ is proved.  

\subsection*{The induction equation on the 3-dimensional torus}

Magnetic fields $\B$ into a compact domain $D$ are naturally generalized
to magnetic fields on the 3-torus.  In this generalization we follows to the results of [C].
Let us denote by  $L_2$ the space of $2\pi$--periodic vector-functions over the standard cube, the square of which is integrable. 
Let us consider the following decomposition into the direct factors: 
$$ L_2 = L_2^0 \oplus L_2^+ \oplus L_2^-, $$
where $L_2^0$ is the space of the kernel of the operator  $\rot$. The spaces 
$L_2^+$ and $L_2^-$ consists of linear combinations of vector-functions $\{ \k e^{\i\k x} \}$, $\k
\in \Z^3 \setminus \{0\}$, if $\k \ne 0$ and of the three constant vector-functions, if $\k=0$; 
in this formula 
$x$ is a 3-dimensional coordinate on the standard cube, $\i$ is the Gaussian unite. The space  $\{
L_2^{\pm}\}$ are the spaces of proper vector-functions of the operator  $\rot$, 
which corresponds to positive, or negative, proper values.

More precisely, the spaces 
 $L_2^+$ and $L_2^-$ admit the following description. Each vector 
$\k \in \Z^3 \setminus \{0\}$ corresponds to the 2-dimensional complex linear space, which is denoted by $P_{\k}$.
The space $P_{\k}$ is represented by the direct sum of the two linear spaces (see the formula (19) in [C]):
\begin{eqnarray}\label{71}
P_{\k} = P_{\k}^+ \oplus P_{\k}^-.
\end{eqnarray}
The vector-functions $\{ \c_{\k}^{\pm} e^{\i \k x} \}$, $\c_{\k}^{\pm} \in
P_k^{\pm}$, are proper of the operator 
$\rot$,  which corresponds to the proper value   $+ \vert \k \vert$, $- \vert \k \vert$ correspondingly. 

In the previous and analogous formula is convenient to replace the lower index 
 $\k$, which corresponds to a wave vector, to a lower index $k= \pm \vert \k \vert$, which corresponds to a proper value.
Then the parameter $k$ is a real discrete parameter, which is considered as the number of the wave vector  
$\k=\k(k)$. The wave vector is a vector from a sphere with the center at the origin of the radius $\vert k \vert$. 
A proper value corresponds to several wave vectors according to the multiplicity. 

The magnetic field $\B$ is decomposed into the series 
\begin{eqnarray}\label{81}
\B = \sum_{k}  (\c_{k}^+ + \c_k^-) e^{\i \k x},
\end{eqnarray}
where  $k$ is the number of the wave vector, which is assumed non-trivial,
 $\c_k^{\pm}$ is a vector from the space  $P_k^{\pm}$, which determines the complex amplitude of the vector-function
with non-vanishing number $k$. The sign of the number $k$ corresponds to the sign of the magnetic helicity.

Propers vector-functions of the operator 
$\rot$ corresponds  the magnetostatic fields. Such a magnetic field corresponds to extrema  of the magnetic energy 
(see [A], section 1, where the first and the second variation of the magnetic energy is calculated). 
For example, the space of wave vectors with the number $k=1$ (up to parallel translations) is 3-dimensional, because the sphere of the radius $1$
contains exactly $6$ unite vectors of the standard  lattice.  Corresponding magnetic fields are called $ABC$-fields (see [A], p. 219).
The dimension of the space of magnetostatic fields as the function of the number of wave vectors is investigated in [C]. 

For a magnetostatic field the solution of the induction equation 
$(\ref{ind})$ is simplified, because of the condition  $\bf{v}=0$ in magnetostatic equilibria states.
The solution of the equation is given by the formula: 
$$ \B(x,t) = \B(x) \exp{(\lambda (t-t_0)(\alpha - \lambda \eta))}. $$
For the considered solution each magnetic line is invariant, Theorem 
 $\ref{th2}$ is simplified. Namely, the evolution of the quadratic helicity is given by the formula:
$$  \chi^{(2)}(t) = \chi^{(2)}(t_0)\exp{(4\lambda (t-t_0)(\alpha - \lambda \eta))}.$$

\section{Ergodic Birkhoff Theorem, configuration space of magnetic lines} 

Define the configurations spaces, which are constructed for magnetic lines of $\B$.
In this section we assume that $\B$ is a magnetic field on the standard 3-sphere 
$S^3$, as in [A],[TGKMSV]. 

Consider the space 
 $S^3 \times \R_1 \times \R_2$, which consists of the triples  $(x,T_1,T_2)$,
$x \in S^3$, $T_1, T_2 \in \R$, this space is called the configuration space of the type  $(1,2)$ and is denoted by  $K_{1,2}$ 
The following mapping  $F: S^3 \times \R_1 \times \R_2 \to S^3 \times S^3$ is well defined by the formula
$$ F(x,T_1,T_2) = (g^{T_1}(x), g^{T_2}(x)), $$
where $g^t$ is the magnetic flow along the magnetic line of $\B$ trough the point $x$ (a pair of points on each magnetic line).

On the tangent space 
 $T(K_{1,2})$ 1-forms $\A_i$, $i=1,2,$ are defined by the formula
$\A_i = p_i \circ F^{\ast}(\A)$, where $p_i: S^3 \times S^3 \to S^3$ is the projection onto the $i$-th factor, 
$\A$ is the 1-form of the potential of  $\B$. On the tangent space  $T(K_{1,2})$ a pair of the 2-forms  $\B_i$, $i=1,2$ are also defined by the formula
$\B_i = p_i \circ F^{\ast}(\B)$ and a pair of the 1-forms $dT_i$ along the corresponding coordinate  $T_i$. 

The formula $(\ref{(2)})$ for the quadratic helicity $\chi^{(2)}$ is rewritten as following:  
$$ \chi^{(2)} = \int  (\A_1, \B_1)(\A_2,\B_2) dK_{1,2}, $$
where $dK_{1,2}= dS^3dT_1dT_2$ is the volume form, and the integration over the 
variables  $T_1,T_2$ is assumed in the asymptotic sense. 

Analogously, the configuration space 
$K_{2,2} = S^3 \times S^3 \times T_{1,1} \times T_{1,2} \times T_{2,1} \times T_{2,2}$ is well defined (a pair of points on each pair of magnetic lines). This space is used to 
rewrite  the integral 
$\chi^{[2]}$.

Using the configuration space  
$K_{1,2}$ let us prove the following theorem.

\begin{theorem}\label{relax}

1. For an almost arbitrary point $x$ there exists the following limit:
\begin{eqnarray}\label{lim}
\lim_{T \to +\infty} \Lambda^{(2)}(T;x) = \lim_{T \to + \infty} \frac{1}{T^2}
(\int_{0}^{T}  (\dot{x}(\tau),\A)d\tau)^2 = \Lambda^{(2)}(x).
\end{eqnarray}
The function  $\Lambda^{(2)}(x)$ is measurable and for almost all magnetic line in the formula  $(\ref{(2)})$ 
the upper limit can be replaced by the limit. The formula itself is rewritten in the following convenient form:  
$$ \chi^{(2)}(\B)=\int \Lambda^{(2)}(x) dx. $$  

2.  The  dispersion of the helicity density $(\A(x),\B(x))$ with the middle value
$$\overline{(\A(x(T)),\B(x(T)))}_{T \ge 0} = \Lambda(x(0))),$$
(which is defined for an almost arbitrary  magnetic line trough an almost arbitrary point $x(0) \in S^3$), 
  coincides after the integration over $S^3$ with the difference 
$$ \delta^{(2)} - \chi^{(2)},$$
where $\delta^{(2)}$ is defined by the formula   $(\ref{delta2})$ (comp. with the inequality  $(\ref{11})$). 

3. The dispersion of the function $\Lambda(x(0))$ with the middle value 
$$\overline{\Lambda(x(0))}_{x_0 \in S^3}$$ 
(which is defined for an almost arbitrary  magnetic line trough an almost arbitrary point $x(0) \in S^3$)
coincides after the integration over $S^3$ with the difference 
$$ \chi^{(2)} - \frac{\chi^2}{Vol S^3} $$
(see the right inequality in the formula  $(\ref{13})$).

4. For Hopf magnetic field, which is defined by the formula  $(\ref{hopf})$, 
the value of the density of the magnetic helicity  $(\A(x),\B(x))$ is a constant. In particular, the inequalities
in Theorem $(\ref{th1})$ are the identity.
\end{theorem}

\subsubsection*{Proof of Theorem  $\ref{relax}$} 

Let us recall Ergodic Birkhoff Theorem. In the book [H] this theorem is proved for discrete flow, which preserve the measure.
The proof is generalized word-by-word to smooth divergent-free one-parameter flows.

\begin{theorem}\label{Bir}
Let  $g^{t}$ be a volume-preserved flow (a one-parameter of volume-preserved diffeomorphisms) on $S^3$ and let $f \in L_1(S^3)$ be an integrable function. Then the mean values   
$f^{\ast}(T,x) = T^{-1}\int_0^T f(g^t(x))dt$ are convergent for almost an arbitrary point   $x \in S^3$, moreover, the limit
$f^{\ast}(x)=\lim_{T \to + \infty} f^{\ast}(T)$ is integrable and is invariant with respect to the flow  $g^{t}$,
namely, $f^{\ast}(g^t(x))=f^{\ast}(x)$. Assuming 
$m(K)<\infty$, we get $\int f^{\ast}(x) dK = \int f(x) dK$. 
\end{theorem}

The flow $g^{t}$ induces the flows $g_{1}$, $g_2$ on the configuration space 
 $K_{1,2}$, which are preserve the measure  $dK_{1,2}$. The flow 
$g_i$, $i=1,2$ transforms  $i$-th point of the pair along the magnetic line. Evidently, the flows $g_1$, $g_2$ are commuted.
Let us apply Theorem 
$\ref{Bir}$ to the flow $g_1$. We get the middle (asymptotic) value along the each $t_1$--trajectory of a point $(x,t_1,t_2)$, 
which is denoted by  $f_1(x,t_2)$. Let us apply twice the Theorem $\ref{Bir}$ to the flow along the each  $t_2$--trajectory of a point
$f_1(x,t_2)$. Because of estimations, which are given by the equation  $(\ref{noneq})$, we get the required result.
The statement 1 is proved. 

Statements 2 and 3 are followed immediately from the identity
$$\int_{S^3}(f(x)-\bar f)^2 dS^3 = \int f^2(x) dS^3 - \bar f^2,$$ 
where
$\bar f = \int f(x)dS^3$, $x \in S^3$.

Statement 4 is evident. Theorem 
 $\ref{relax}$ is proved.

\section{A possible role of higher helicities during relaxations of magnetic fields}

In the paper by V.I.Arnol'd [A] the space of magnetic fields on a closed 3-dimensional manifold
 $M^3$ is studied (Example  5.2 p. 234). In the case the manifold $M^3$ coincides with the standard sphere  $S^3$, 
 the global minimum of the magnetic energy corresponds to Hopf magnetic field.
This magnetic field is defined by the 2-form 
\begin{eqnarray}\label{hopf}
h^{\ast}(d\omega), 
\end{eqnarray}
where $h: S^3 \to S^2$ is the standard Hopf fibration,  $d\omega$ is the standard area  form on $S^2$.

Theorem 
 $\ref{relax}$ clarifies a possible role of higher helicities invariants in MHD.
It is well known (see [T]) that during a relaxation of magnetic fields to a magnetostatic field  with minimal magnetic energy
a recconection of magnetic lines is possible. This recconection keeps the helicity invariant. 
By Theorem $\ref{th2}$ the quadratic helisity $\chi^{(2)}$ is continuous. Moreover, the absolute value of the velocity of fluctuations of the quadratic magnetic helicity 
is estimated from below. 

The quadratic helicity is the simplest topological invariant, which is measured the dispersion of the helicity density $(\A,\B)$
from its middle value $\chi(\B)$. During a relaxation of magnetic field the magnetic energy is decreased to the minimal possible value,
which is prescribed by the V.I.Arnol'd inequality (see the formula $(\ref{Arnold})$ below), and the dispersion of the helicity fluctuation is minimized.
(see the statements 2,3,4 of Lemma $\ref{relax}$).

An analogous meaning has the quadratic helicity 
$\chi^{[2]}$, and polynomial helicities, which are related with higher momenta of the helicity density.
In particular, polynomial helicities can be used to investigate small magnetic dissipations of magnetostatic fields. 

The higher helicity, which is defined in [Akh], is related with the simplest polynomial invariants of the magnetic knots, which is not 
expressed from the (asymptotic) linking coefficients.  
What did the role played such higher helicities  during relaxation of magnetic fields? 

A non-twisted magnetic tube, which is self-linked in the form of the "`trefoil"' knot, consists of the family of magnetic lines, which are pairwise linked with the coefficient $+3$ (see, f.ex. the figure 
6.8 [Z-R-S, ��.6]). In particular, this means that for the considered magnetic field the difference  
$ \chi^{(2)} - \frac{\chi^2}{Vol S^3}$ is equal to zero (there is no fluctuation of the helicity density on the spectrum of magnetic lines). 
It is well-known that for the considered  magnetic field could be assumed a magnetostatic field with a local minimum of the magnetic energy. 
After a sufficiently  perturbation of the magnetic field a reconnection of magnetic lines are observed. 
As the result of this reconnection the magnetic field is transformed to the Hopf magnetic field with the global minimum of the magnetic energy.
The magnetic helicity is kept during the reconnection.

During the reconnection inside the knotted magnetic tube a fluctuation of the magnetic helicity is appeared: pairwise linking numbers
of magnetic lines are variated, in particular this linking numbers could be different because of a non-homogeneous twisting of the knotted magnetic tube  along the transversal cross-sections to the central line. 
For such a tube the higher magnetic helicity, generally speaking, is non-trivial.  
The higher helicities characterize knotting of magnetic tubes in the case of the fluctuation of the helicity density on the spectrum of magnetic lines,
in particular, when the difference  
$ \chi^{(2)} - \frac{\chi^2}{Vol S^3}$ is positive.

An example of higher helicity is given by the helisity of a family of special 2-forms, which is constructed using electric component of the
field on the configuration space of magnetic lines. In this example the higher magnetic helicity is a skew-invariant with 
respect to the mirror symmetry of the magnetic field and has the dimension 
 ��$^{12}$��$^{6}$.

On the configuration space, which is associated with all triples of magnetic lines trough points  
 $x_1,x_2,x_3$ on the sphere $S^3$, the following  1-forms $(2,3)\A(x_1)$, $(3,1)\A(x_2)$, $(1,2)\A(x_3)$, 
which are defined using electric currents through magnetic lines are well-defined.  In this construction 
$\A(x_i)$ is the vector-potential of the corresponding magnetic line at the point $x_i$, $(i,j)$
is the (asymptotic) linking number of the magnetic line at the points
 $x_i$, $x_j$. The considered 1-forms determine the following  2-form 
$$F(x_1,x_2,x_3)=(1,3)(3,2)\A(x_1) \times \A(x_2) + $$
$$(2,1)(1,3)\A(x_2) \times \A(x_3) + (3,2)(2,1)\A(x_3) \times \A(x_1),$$ 
from $\Lambda^2(T(S^3))$, which is exact in a suitable gauge. 
The integral mean value over the configuration space of all triple of magnetic lines of the helicities of the considered family of 2-forms
is called the higher helicity and is denoted by 
 $M$.  (The helicity of the electric component of the magnetic field was studied before f.ex. in 
 [R-S-T]).

During a relaxation of the magnetic field to the Hopf magnetostatic field with the minimum of the magnetic energy the higher magnetic
helicity is destroyed and, for magnetic fields with no knotted tubes, even with a fluctuation of the magnetic helicity, is equal to zero. 
This corresponds to a simplification of a complicated topological configuration of magnetic fields
during a magnetic reconnection.

\section{A remark about the Arnol'd inequality}

 The Arnol'd inequality is following. For a magnetic field in a domain $D$ there exists a constant $C>0$, 
which is depended only on geometric properties of the domain, such that the following inequality is satisfied: 
\begin{eqnarray}\label{Arnold}
C^{-2}U^2(\B) \ge \chi^2(\B).
\end{eqnarray}
For magnetic fields on the 3-torus this inequality also is satisfied and can be proved by means of an estimation of the proper values
of the operator 
 $\rot$, using the spectral decomposition of the magnetic field $\B$ (see section 3). 

A large-scale magnetic fields exist with small-scale magnetic fields.
In the book [Z-R-S, chapter 8, section III] the authors explain how to estimate small-scale magnetic fields.
The coefficients of the spectral decomposition of  $\delta^{(2)}$, 
  which are easily to calculate and the coefficients of  $\chi$
give additional information about the spectra of the magnetic energy $U(\B)$. 
Let us consider as an example the following calculation.

Let us assume that the spectra of a small-scale magnetic field is polynomial. Using this assumption the magnetic energy is given by the following
series:
$$ E = \sum_{k} \c_k^+ \bar{ \c}_k^+ + \c_k^- \bar{ \c}^-_k = \sum_k \vert \c_k^+ \vert^2 + \vert \c_k^- \vert^2, $$
assume as an example that  
$$ c^{\pm}_k = \gamma^{\pm} k^{-\alpha}, \quad \alpha = \frac{5}{3}, \quad c_k^{\pm} = \vert \c_k^{\pm}
\vert. $$

The square of the magnetic energy is given by the following series:
\begin{eqnarray}\label{91}
E^2 = \sum_{k} \frac{2(\gamma^+ + \gamma^-)^2}{\alpha - 1}
k^{-2\alpha + 1}.
\end{eqnarray}

In this formula the terms are calculated as follows.
Let us replace in the formula  $(\ref{81})$ the discrete parameter  $k$ to the continuous parameter 
$k \ge 1$, which is denoted the same. 
In the formula $(\ref{91})$ the spectral distribution is defined by the integration over the $k$-plane,
the first coordinate sector of the radius    
 $k^2_1 + k_2^2 = k$ is replaced by the square  $k_1
\le k$, $k_2 \le k$.

The helicity of the magnetic field
 $\B$ is decomposed into the following series: 
$$ \chi = \sum_{k} b^+_{k} - b_k^-, $$
where the coefficients 
 $b^{\pm}_k$ are non-negative. 

Analogous calculations give
$$ b_k = b_k^+ - b_k^- = (\gamma^+ - \gamma^-) k^{-\alpha-1}.$$
The square of the magnetic helicity is decomposed as follows:
$$ \chi^2 = \sum_{k} b^{(2)}_{k}. $$
Evidently, we get:
$$ b^{(2)}_k = \frac{2 (\gamma^+ - \gamma_-)}{\alpha} k^{-2\alpha-1}.$$

The correlation tensor 
 $\delta^{(2)}$ is decomposed as follows: 
$$ \delta^{(2)} = \sum_k d^{(2)}_k. $$
The upper bound is given by the following expression:
$$ d^{(2)}_k \le \frac{\gamma^+ + \gamma^-}{\alpha^2} k^{-2\alpha}.$$

The last calculation is simple. All the quadruples of wave vectors 
 $\k_1, \k_2, \k_3, \k_4$, which are satisfied the condition
\begin{eqnarray}\label{10}
\k_1 + \k_2 = \k_3 + \k_4
\end{eqnarray}
has to be calculated. 
For the magnetic field with the wave vectors
$\k_1$, $\k_2$ (correspondingly, with the wave vectors $\k_3$, $\k_4$) the magnetic field with the wave vector  $\k_1 + \k_2$ 
(correspondingly, with the wave vector $\k_3 + \k_4$ 
is standardly calculated. 
The corresponding vector-potential is defined by addition $-2$ to the corresponding spectral index. 
A non-trivial convolution is possible if the condition 
$(\ref{10})$ is satisfied.  The spectra of $U^2$ and $\chi^2$ are given by the 2-dimensional convolution,
the spectrum of  
$\delta^{(2)}$ is given by the 3-dimensional convolution. This proves that the spectrum
of $\delta^{(2)}$ (and of 
$\chi^{(2)}$) is between the spectra of $U^2$ and $\chi^2$.

\end{document}